\documentclass[12pt,thmsa,a4paper]{article}
\usepackage{sw20lart}

\input tcilatex
\begin{document}

\author{M. Enkelmann, U. Werthenbach, G. Zech, T. Zeuner \\
{\small University of Siegen, Department of Physics, 57068 Siegen, Germany}}
\title{An optical readout for a fiber tracker }
\maketitle

\begin{abstract}
The performance of 16 and 64 channel photomultipliers coupled to
scintillating fibers has been tested. The devices are sensitive to single
photoelectrons, show little gain losses for magnetic fields up to 100 Gauss
and have moderate optical cross-talk. The maximum channel to channel gain
variations reach a factor two for the 16 channel version and a factor of
four for the 64 channel PM. The measurements and simulations indicate that
the photomultipliers are well suited for the light detection in fiber
trackers.
\end{abstract}

\section{Introduction}

Scintillating fiber trackers constitute an interesting alternative to
gaseous tracking detectors. Impressive progress has been made in the last
years in both the quality of plastic fibers and the optical readout \cite
{leu95}\cite{dam96}\cite{adam95}. Fiber detectors avoid all the problems
related to HV connections, sparks, electronic noise and pick-up in wire
chambers and MSGC detectors. In view of these advantages the Hera-B \cite
{herab} collaboration investigated the possibility to build an inner tracker
of scintillating fibers as an alternative to a MSGC system. The fiber
solution was finally abandoned due to the higher cost for a 150000 channel
system and expected radiation damage. Nevertheless the results of the study
are of possible interest for detectors operating under different
experimental conditions.

Sensitive light detection is most efficiently achieved with
photomultipliers. To reduce cost and space one has to use multichannel
systems. Interesting alternative solutions provide multichannel PMs (MCPM)%
\cite{leu95}\cite{dam94} built in standard technique, hybrid PMs (HPMT)\cite
{dam94a}\ and visible light counters (VLPC)\cite{adam95}. The latter provide
by far the best photoefficiency of up to 80 \%, but need cooling to below
10K. HPMTs allow for a finer spatial segmentation than MCPMs but require
higher voltages and low noise amplification. In this article we present
measurements with a readout system consisting of fibers coupled to small
size 16 and 64 channel photomultipliers \cite{yos97}.

In the following sections we describe the setup, the simulation of the
photon propagation and, finally, the measurements which concentrated on the
investigation of the influence of magnetic fields and the cross-talk between
channels.

\section{The setup}

\FRAME{ftbpFU}{3.5258in}{2.7562in}{0pt}{\Qcb{Setup to measure the optical
cross-talk. }}{}{aufbau.ps}{\special{language "Scientific Word";type
"GRAPHIC";display "PICT";valid_file "F";width 3.5258in;height 2.7562in;depth
0pt;original-width 568.9375pt;original-height 818.8125pt;cropleft
"0.0436875";croptop "0.7954013";cropright "0.9663484";cropbottom
"0.2158048";filename 'C: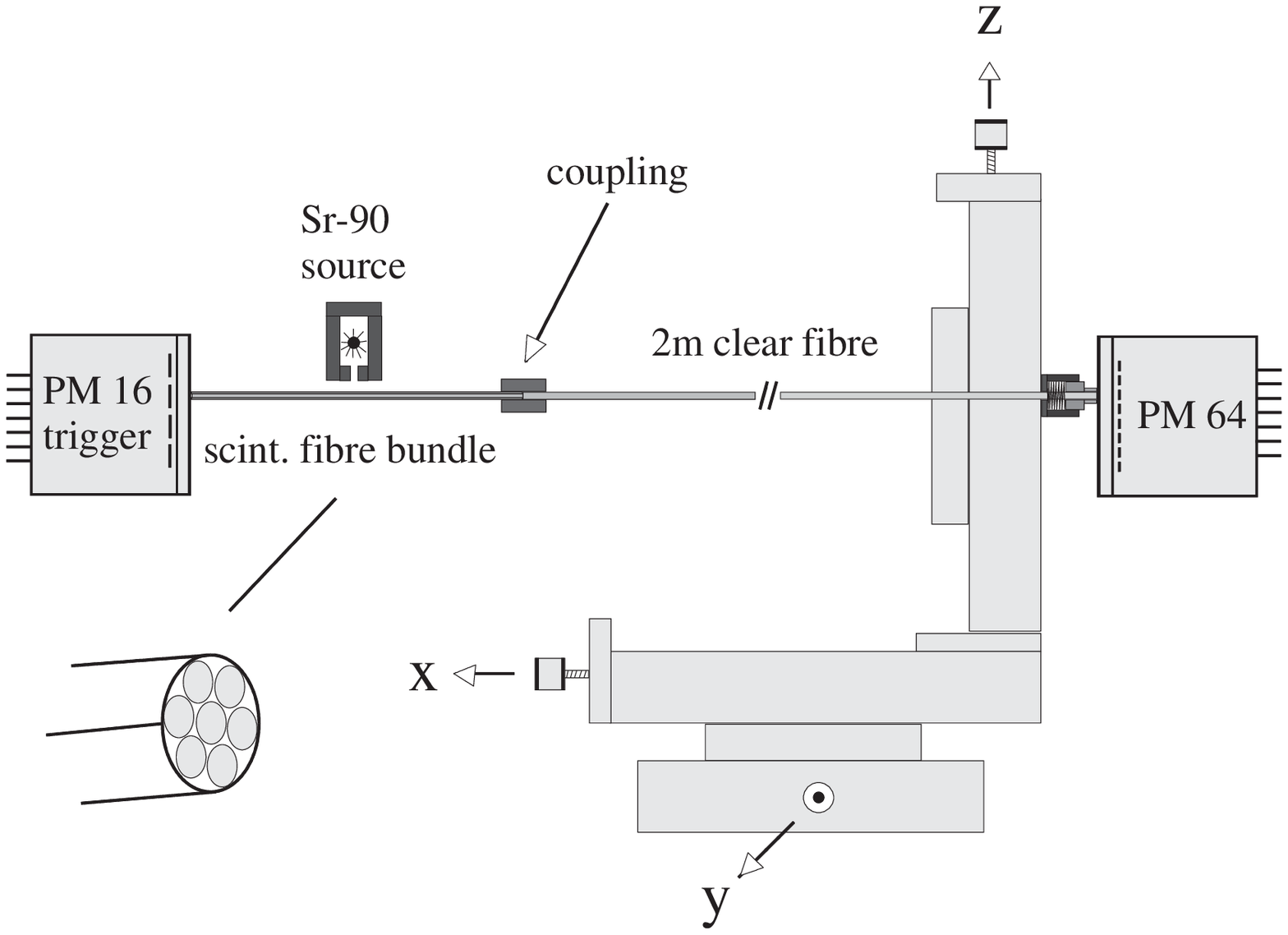';file-properties "XNPEU";}}We
investigated multichannel photomultiplier tubes\footnote{%
HAMAMATSU Photonics K.K., Japan} \cite{yos97} with 16 (M16) and 64 channels
(M64) on a square sensitive area of $18\times 18mm^2$. The tubes have
bialkali photocathodes and metal channel dynode structures of front pad
sizes of $4\times 4mm^2$ (M16) and $2\times 2mm^2$ (M64). The gaps between
the dynodes of the matrix next to the window are $0.5mm$ (M16) and $0.3mm$
(M64) respectively.

Two different setups were used to measure the gain uniformity of the PM
tubes and the optical cross-talk between adjacent channels. For the
uniformity measurement a LED was operated at a distance of $30cm$ in front
of the PM tube. The experimental arrangement for the cross-talk measurement
(see Fig.1) consisted of a fiber guide mounted on a x-y-z-table which was
moved under computer control across the PM window. The fiber was pressed by
a spring onto the PM window.

In order to investigate the PM performance in magnetic fields, the PM was
placed between two Helmholtz coils providing fields of up to $130$ Gauss.

For the uniformity measurement the light was produced by standard green or
blue LEDs. Short light pulses of the LED were generated by a special high
current control circuit with adjustable pulse width. The optical cross-talk
measurements were performed with scintillation light produced by $\beta ^{-}$
rays from a Sr$^{90}$-source in scintillating fibers.

\FRAME{ftbpFU}{2.0124in}{3.8164in}{0pt}{\Qcb{Active high voltage divider
chain.}}{}{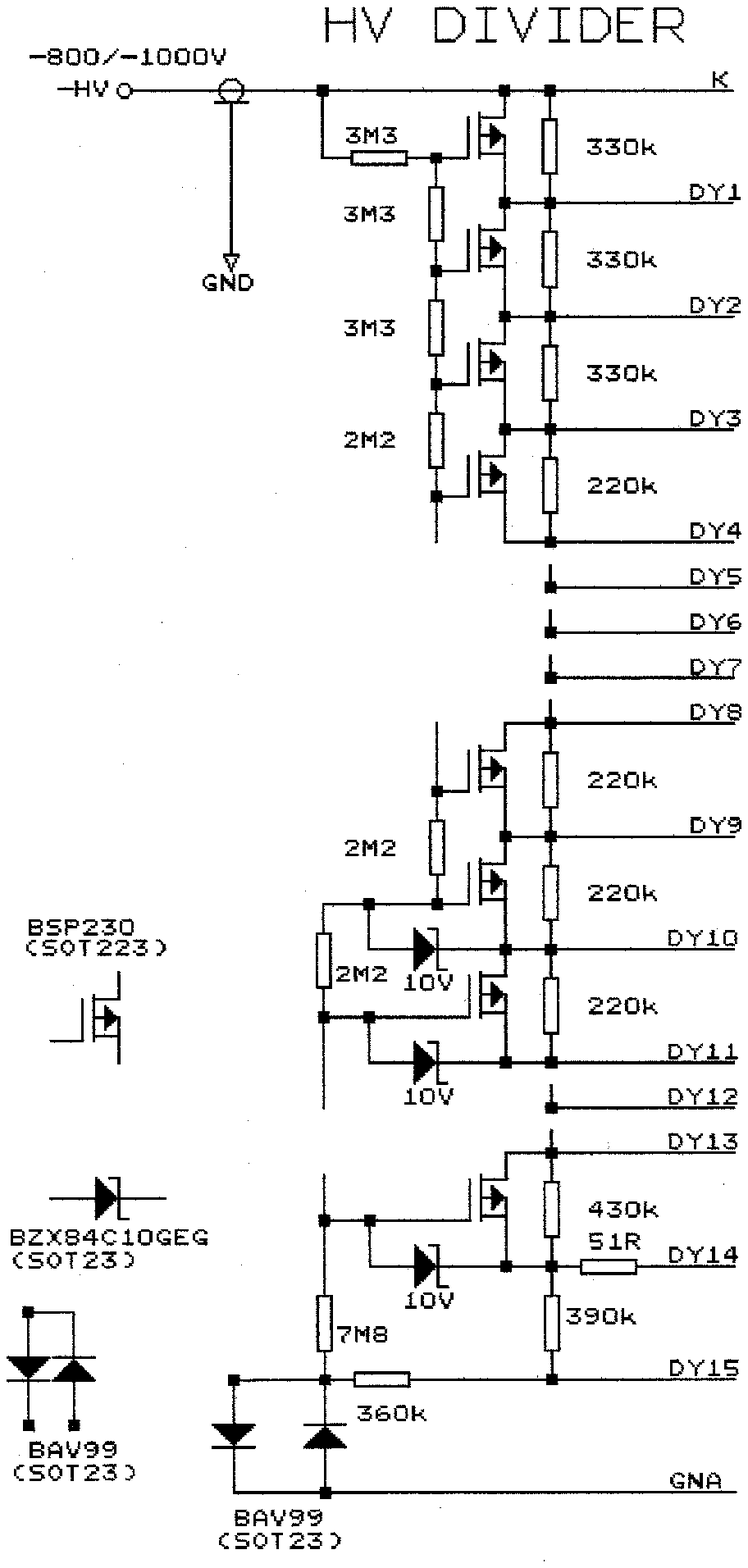}{\special{language "Scientific Word";type
"GRAPHIC";maintain-aspect-ratio TRUE;display "USEDEF";valid_file "F";width
2.0124in;height 3.8164in;depth 0pt;original-width 305.5625pt;original-height
583.125pt;cropleft "0";croptop "1";cropright "1";cropbottom "0";filename
'C:/pmpaper/schalt.eps';file-properties "XNPEU";}} The anode signals of the
PM channels were amplified by standard NIM linear amplifiers with a fixed
gain factor of $10$. They were digitized by CAMAC ADCs coupled to a PC.

In the Hera-B experiment rates of $200kHz$ per channel are expected with
average signal of 20 photoelectrons. The total photocurrent reaches $0.1mA$
for the 64 channel tube. To reduce the bleeder current an active voltage
divider circuit for the dynodes (Fig. 2) was developed. Compared to a
passive chain the current and correspondingly the number of HV power
supplies and the heat dissipation can be reduced by a factor of five.

\section{Simulation of photon propagation}

Photons produced in a scintillating fiber are usually transferred to the
photosensitive device via clear plastic fibers. The coupling of the fibers,
reflection losses and the connection of the clear fiber to the PM affect the
photoelectron yield and their lateral distribution. We have simulated these
processes. For the numerical estimates we used refraction indices for the
fiber ($n_f$), its gladding ($n_{cl}$), the PM window ($n_w$) and the
photocathode ($n_{ca}$) of $n_f=1.60$, $n_{cl}=1.42$, $n_w=1.53$ and $%
n_{ca}=3.4$. \FRAME{ftbpFU}{3.525in}{2.6852in}{0pt}{\Qcb{Trapping efficiency
across a scintillating fiber. Spiraling photons are lost when the
scintillating fiber is coupled at the center of a clear fiber.}}{}{fibef.ps}{%
\special{language "Scientific Word";type "GRAPHIC";display "PICT";valid_file
"F";width 3.525in;height 2.6852in;depth 0pt;original-width
568.9375pt;original-height 818.8125pt;cropleft "0.0223125";croptop
"0.7359812";cropright "1.0061639";cropbottom "0.2788293";filename
'C:/pmpaper/FIBEF.ps';file-properties "XNPEU";}}

In the following we assume that the cross section of the fibers is circular.
The trapping probability of the photons then depends on the distance of the
production point from the surface. The trapping efficiency across the fiber
is shown in Figure 3. The high efficiency for photons produced near the
surface is due to the trapping of photons with rather large angles with
respect to the fiber axis which then spiral along the surface in azimutal
direction. The average trapping efficiency is $t=1-(n_{cl}/n_f)^2=0.213$.
When the scintillating fiber is coupled to a clear fiber a large fraction of
the spiraling photons is lost, the exact amount depending on the ratio of
the diameters and the relative lateral positions of the axises. In the limit
where photons are produced uniformly in the scintillating fiber, where the
scintillating fiber is coupled axially at the center of the clear fiber and
where the radius of the scintillating fiber is negligible compared to the
clear fiber radius the trapping efficiency is reduced by a factor of about
two: $t=1-n_{cl}/n_f=0.113$.

Spiraling photons suffer from a high number of reflections at the fiber
surface and thus have a higher probability to be lost than photons crossing
the fiber axis. The losses depend on the quality of the surface and on the
circular symmetry of the fiber. Often the cross section of so-called round
fibers is in reality slightly elliptical and varying along the fiber. These
losses which are difficult to estimate have also a positive aspect: They
reduce the light divergence at the exit of the fiber.

For an efficient detection of minimum ionization particles in scintillating
fibers with PMs of about $10\%$ conversion efficiency a minimum scintillator
thickness of about 1 to 2 mm is required, the exact value depending on the
quality of the fiber and many other parameters related to the light
detection system. Usually several thin scintillating fibers have to be
coupled to one clear fiber to obtain an efficient and precise tracking. The
minimum radius of the clear fiber which one chooses as small as possible is
determined by the number and the cross section of the scintillating fibers
to be coupled. On the other hand the maximum allowable cross section depends
on the size of the PM channels, the window thickness and the light
divergence at the fiber exit.\FRAME{ftbpFU}{3.6685in}{2.5538in}{0pt}{\Qcb{%
Photon loss due to non-parallel coupling of the fibers. The solid curve
represents the worst case, where the scintillating fiber is coupled at the
center of a clear fiber. The upper curves correspond to fibers coupled at
the periphery.}}{}{tilt.ps}{\special{language "Scientific Word";type
"GRAPHIC";display "PICT";valid_file "F";width 3.6685in;height 2.5538in;depth
0pt;original-width 568.9375pt;original-height 818.8125pt;cropleft
"0.0221525";croptop "0.5326729";cropright "0.6886484";cropbottom
"0.1806249";filename 'C:/pmpaper/TILT.ps';file-properties "XNPEU";}}

Figure 4 shows the effect of a tilt of the two axises in the coupling of the
fibers. Losses start to become important only for rather large tilt angles.

The cross-talk to neighbouring PM channels is illustrated in Figure 5. The
curves correspond to the fraction of photoelectrons collected by neigbouring
channels surrounding one of the central pads. (Of course this fraction is
lower for channels at the borders or corners of the PM.) The photons are
produced in a bundle of seven scintillating fibers ($0.5mm$ diameter)
coupled to one clear fiber of $3m$ length. The cross-talk becomes disturbing
when the fiber diameter approaches the size of the square channel pads and
is especially important for the standard window of $1.3mm$ thickness. For a
signal of in average 10 photoelectrons and single photoelectron detection%
\FRAME{ftbpFU}{4.4278in}{3.5224in}{0pt}{\Qcb{Simulation of optical
cross-talk for 64 channel PM as a function of the fiber radius.}}{}{%
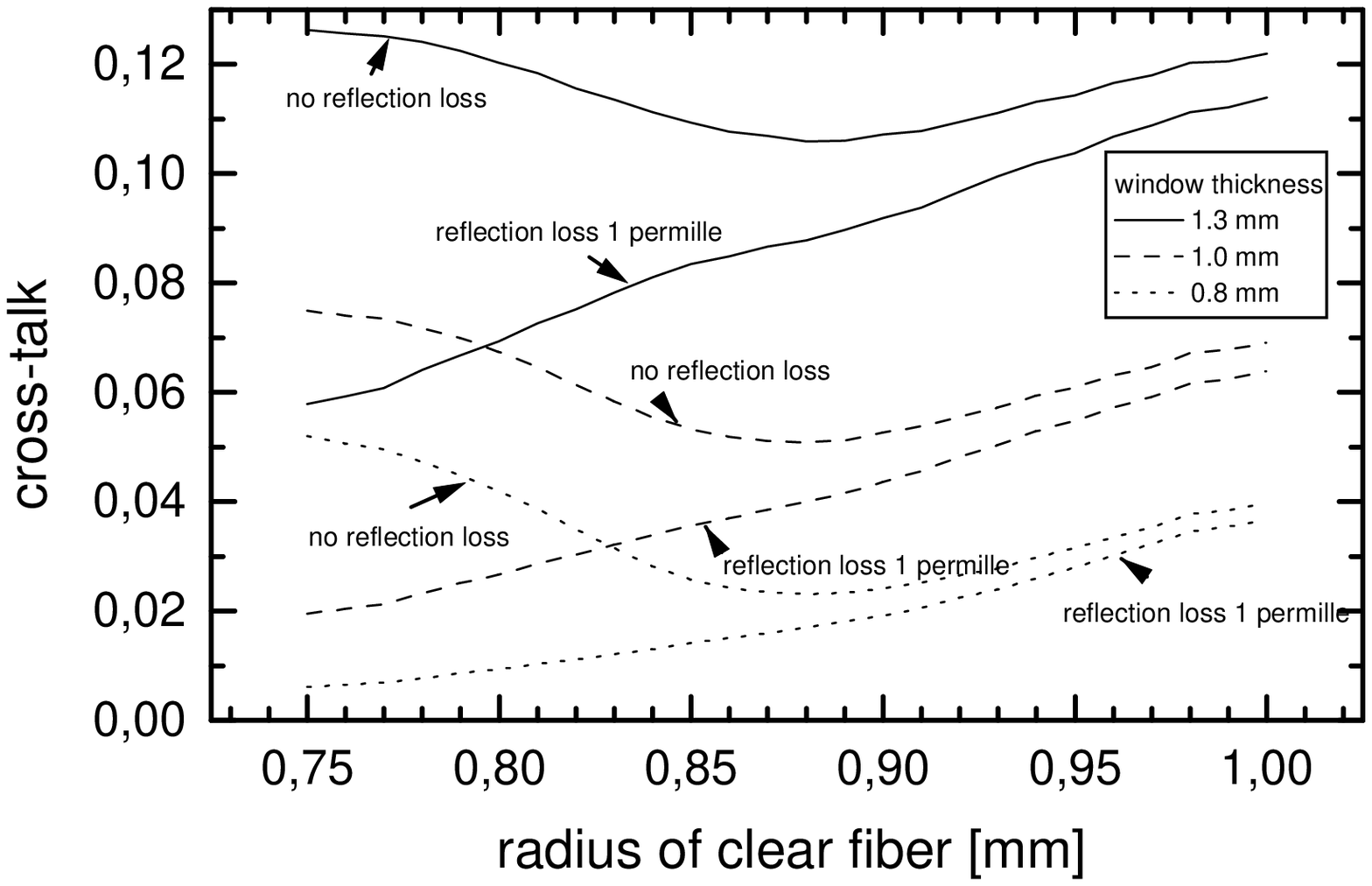}{\special{language "Scientific Word";type "GRAPHIC";display
"USEDEF";valid_file "F";width 4.4278in;height 3.5224in;depth
0pt;original-width 568.9375pt;original-height 818.8125pt;cropleft
"0.0183592";croptop "0.764805";cropright "0.86705";cropbottom
"0.2969302";filename 'C:/pmpaper/CTSIMULA.ps';file-properties "XNPEU";}} a
channel multiplicity between $1.6$ and $2.2$ for a $1.5mm$ diameter fiber is
expected.

The large cross-talk for small radia and negligible reflection loss is due
to the increase of the number of spiraling photons when the coupled
scintillating fibers cover the full cross section of the clear fiber. If
sufficient light is available, the cross-talk can be reduced drastically by
setting the threshold above the one photoelectron signal.

\section{Measurements and results}

We have tested six M16 tubes and three 64 channel prototypes with window
thicknesses of $1.3mm$, $1.0mm$ and $0.8mm$.

\subsection{Signals, amplification}

The photomultiplier tubes have a fast time response. For the 16 channel
device the pulse width for a single photoelectron is of the order of $1$ to $%
2ns$, the transit time spread is $0.3ns$ and the gain at the nominal voltage
of $800V$ is $3\cdot 10^6$ $^2$. The numbers for the 64 channel tube are
very similar except for a lower gain of $3\cdot 10^5$ $\footnote{%
HAMAMATSU Product specifications.}.$

For optical fiber read-out in fiber trackers the photomultipliers have to
work with very low light levels. Efficient detection of single
photoelectrons is therefore required. Figure 6 shows pulse height
distributions for single photoelectrons measured at different high voltage
settings for the M16 type. At voltages above $900V$ the single photoelectron
signal is clearly separated from noise with a signal to noise ratio larger
than ten$.$ The M64 type has a smaller gain than the 16 channel version but
a detection of single photoelectrons is also possible with an adapted
readout system.\FRAME{ftbpFU}{4.28in}{3.5155in}{0pt}{\Qcb{Single
photoelectron pulse height distribution for 16 channel PM for different HV
settings.}}{}{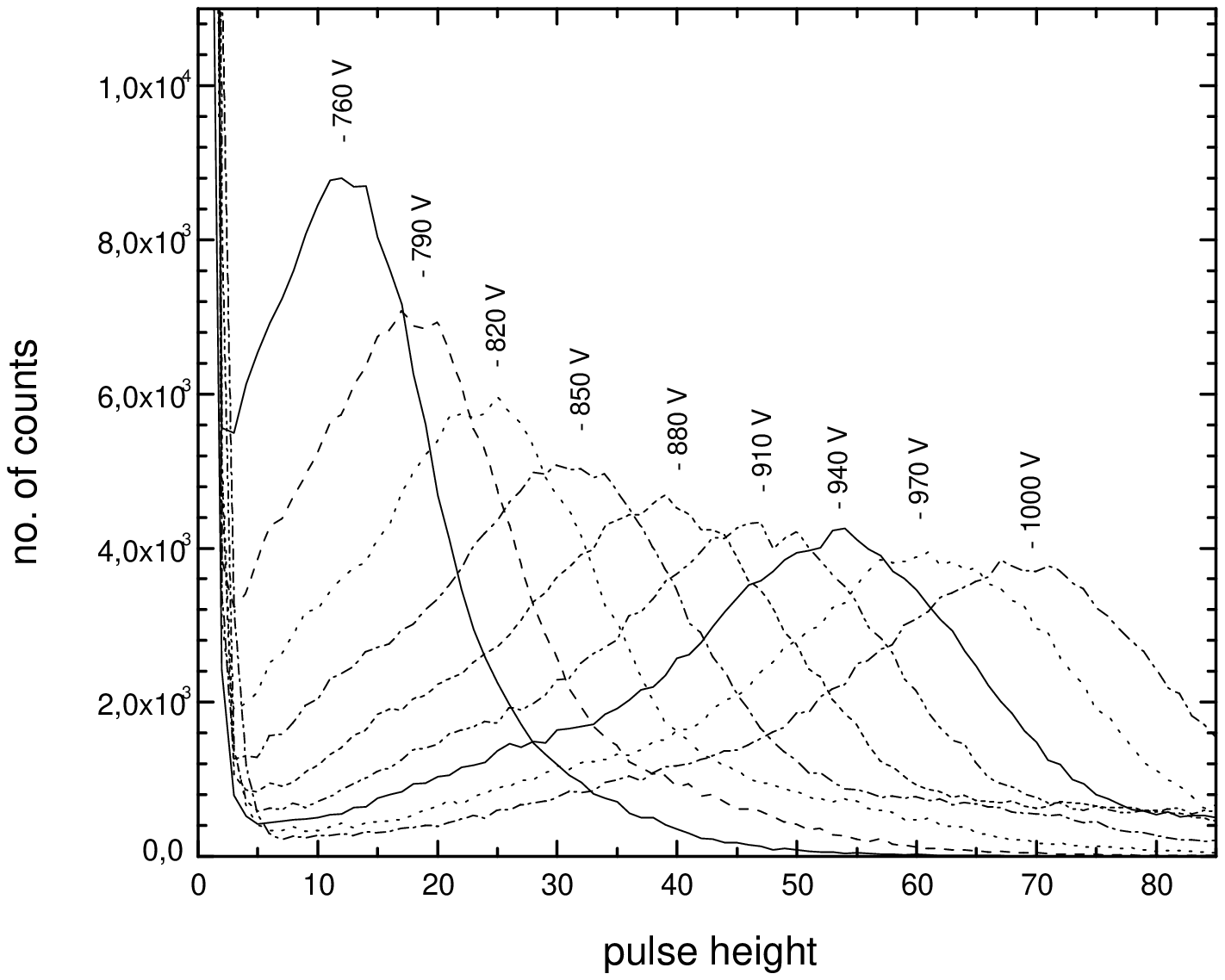}{\special{language "Scientific Word";type
"GRAPHIC";display "PICT";valid_file "F";width 4.28in;height 3.5155in;depth
0pt;original-width 568.9375pt;original-height 818.8125pt;cropleft
"0.063629";croptop "0.900244";cropright "0.841535";cropbottom
"0.435104";filename 'C:/pmpaper/HV6.ps';file-properties "XNPEU";}}

\subsection{Uniformity}

The channel to channel gain variations of the M16 and M64 photomultipliers
were measured using a LED at a distance of $30cm$ from the PM which
illuminated the whole photocathode. This configuration generated single
photoelectron signals. All channels were read-out in parallel. The relative
gain was calculated from the single photoelectron peak position in the pulse
height distribution. The maximum gain variation between different channels
of the same tube is less than a factor of two for the M16 PM's , for M64
PM's the variation is larger reaching a factor of four. The gain uniformity
was measured for 6 PM's of the M16 type and the first two available
prototypes of the M64 series. Figure 7 shows for both cases a typical gain
histogram.

The mean gain variation of the six tested photomultipliers of the M16 type
is $25\%$.\FRAME{ftbpFU}{4.0577in}{5.9845in}{0pt}{\Qcb{Uniformity of the
response of the channels for the 64 and the 16 channel PMs.}}{}{hom64_16.ps}{%
\special{language "Scientific Word";type "GRAPHIC";maintain-aspect-ratio
TRUE;display "PICT";valid_file "F";width 4.0577in;height 5.9845in;depth
0pt;original-width 568.9375pt;original-height 818.8125pt;cropleft
"0.006314";croptop "1.032848";cropright "1.04174";cropbottom
"-0.006119";filename 'C:/pmpaper/HOM64_16.ps';file-properties "XNPEU";}}

\subsection{Position sensitivity and cross-talk}

We have measured the position sensitivity of the PMs between the individual
channels by scanning across three pads with a thin fiber of $50\mu m$
diameter, where light was injected by a LED parallel to the axis. The
response is rather uniform except for the gap region where the efficiency is
reduced to an average of 60\%. Thus the effective inefficient region is $%
0.2mm$ for the M16 and $0.1mm$ for the M64 tube.

In fiber tracker applications cross-talk between adjacent channels increases
the apparent occupancy and produces fake signals. There are two possible
sources for cross-talk: optical and electrical coupling.

Electrical cross-talk between adjacent dynode structures might occur during
the electron multiplication processes. Up to the level of about ten
photoelectrons electrical cross-talk was not observed in our measurements
indicating that for applications with low light levels the electrical
cross-talk is negligible. This is explained by path length variations and
fluctuations in the light emission time, coincidences between several
photoelectrons within the short transit time are unlikely in most cases.

The optical cross-talk was studied in a similar way as the sensitivity by
moving a fiber of diameters of $1.5mm$ in steps of 100 $\mu m$ across three
dynode pads. However, the light was produced by irradiating a fiber bundle
of seven scintillating fibers with a Sr$^{90}$-source to achieve a more
realistic light distribution. The fiber bundle was coupled to a $2m$ long
clear fiber which was connected to the PM. When the fiber is centered at a
pad the neigbouring channel receives a fraction of $0.023$ of the light
(Fig. 8). Thus for four adjacent channels and a ten photoelectron average
signal the channel multiplicity will be about two for a single
photoelectrons threshold. This effect is especially disturbing since
adjacent PM channels will not always correspond to adjacent fibers in the
tracking detector.

\FRAME{ftbpFU}{4.35in}{3.1842in}{0pt}{\Qcb{Relative photon yield as a
function of the fiber position for two different window thicknesses.
Adjacent channels detect $2.3\%$ ($1.3mm$ window) and $0.9\%$ ($0.8mm$
window).}}{}{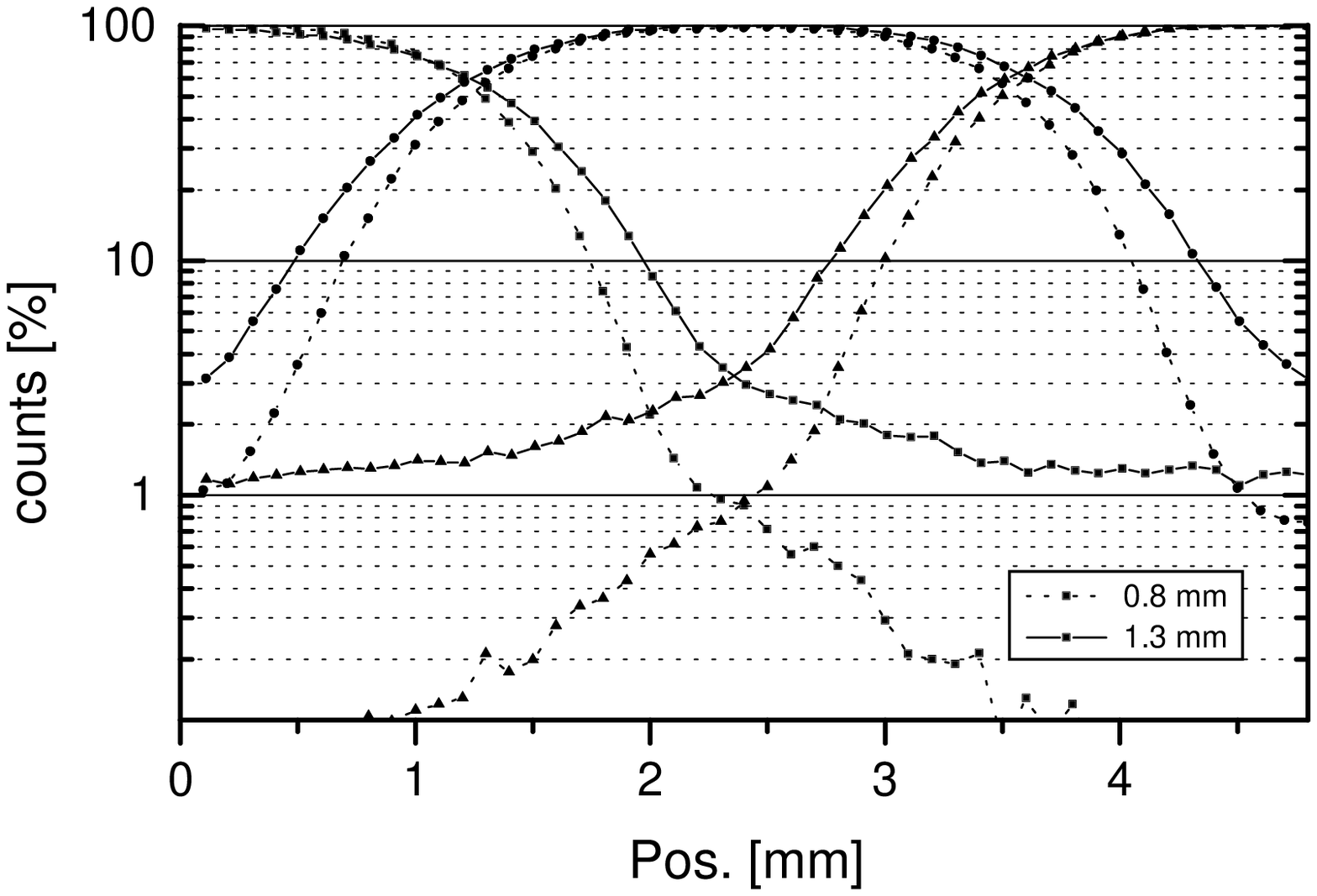}{\special{language "Scientific Word";type
"GRAPHIC";display "PICT";valid_file "F";width 4.35in;height 3.1842in;depth
0pt;original-width 568.9375pt;original-height 818.8125pt;cropleft
"0.024887";croptop "0.754395";cropright "0.884614";cropbottom
"0.302749";filename 'C:/pmpaper/VGL13_08.ps';file-properties "XNPEU";}}

To reduce the optical cross-talk we asked the PM supplier\ to decrease the
entrance window thickness of the photomultiplier tubes to the minimum
thickness which is technically possible. We received two prototype tubes
with entrance window thicknesses of $1.0mm$ and $0.8mm$. The cross-talk was
reduced by a factor $2.5$ for the $0.8mm$ thick window.

Our simulations (Fig. 5) predict for the two cases cross-talks of $0.05$ ($%
0.8mm$ window) and $0.13$ ($1.3mm$ window) for four adjacent channels and
negligible reflection losses and correspondingly $0.005$ and $0.06$ when we
assume that we loose a fraction of $0.001$ for each photon reflection. The
measurements ($0.035$) and ($0.09$) are located in between these predictions.

\subsection{Behaviour in a magnetic field}

At Hera-B the photomultipliers would have to be operated in regions with
magnetic fields of several Gauss up to one Tesla. The gain of conventional
photomultipliers is strongly affected by magnetic fields. As a result of the
deflection of the electrons by the magnetic field in MCPMs in addition to
reduced gain increased cross-talk might be expected.\FRAME{ftbpFU}{4.3734in}{%
3.186in}{0pt}{\Qcb{PM signal as a function of the magnetic field strenth.}}{%
}{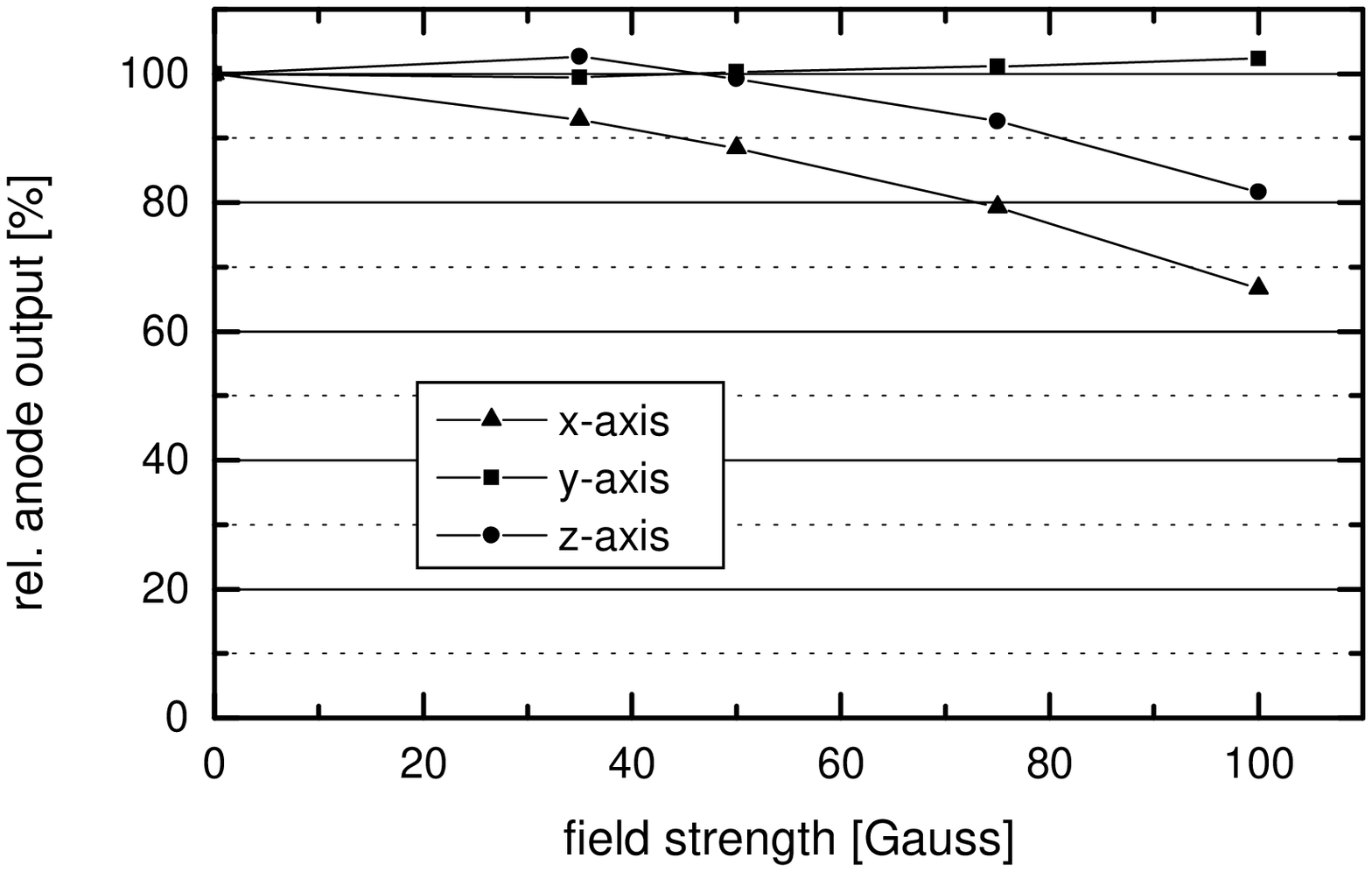}{\special{language "Scientific Word";type "GRAPHIC";display
"PICT";valid_file "F";width 4.3734in;height 3.186in;depth 0pt;original-width
568.9375pt;original-height 818.8125pt;cropleft "0.0363237";croptop
"0.694499";cropright "0.9105767";cropbottom "0.258447";filename
'C:/pmpaper/MAGRST.ps';file-properties "XNPEU";}}

The influence of a magnetic field on the PM performance was measured with
fields up to $100G$ in all three directions with respect to the PM. We used
single photoelectron signals. The magnetic field behaviour of the M16 and
M64 types is very similar as expected because both tubes have the same metal
channel dynode structure except for different lateral dimensions. Figure 9
shows the relative anode current for the 64 channel PM as a function of the
magnetic field in the x-, y- and z-directions where the z axis coincides
with the PM axis. The PMs are rather insensitive to magnetic fields. Within
the precision of our measurement there is no effect for fields parallel to
the y-axis. For the two other field orientations the PM signal is reduced to 
$80\%$ (z-axis) and $70\%$ (x-axis). About half of the loss is due to a gain
reduction, the remaining fraction is loss in efficiency probably due to the
deflection of the first photoelectron.

A position scan with and without magnetic field showed no difference in the
amount of cross-talk. Thus the MCPMs can well be operated in magnetic fields
below $100G$.

\section{Conclusions}

We have studied the possibility to use multichannel photomultipliers to
read-out scintillating fiber trackers. The device is well suited for this
purpose. It is fast, sensitive to single photoelectrons and can be operated
in magnetic fields up to about $0.01$ T. No electric cross-talk was
observed. The optical cross-talk is of the order of $10\%$ for the $1.3mm$
version and single photoelectron detection for fiber diameters near the pad
dimensions. It was reduced by a factor $2.5$ by replacing the $1.3mm$ window
by a $0.8mm$ version. The measured cross-talk is compatible with the
simulations which also show that it depends strongly on the amount of
reflection losses in the fibers. The gain variations from channel to channel
are rather large for the 64 channel tube and require individual off-line
corrections for applications where the analog signal has to be recorded.

\textbf{Acknowledgement}

We thank Drs. Y. Yoshizawa and J. Takeuchi for lending to us the
photomultiplier prototypes and for providing us with valuable informations.
We are grateful to our ingeneers W. U. Otto and R. Seibert for the
development of high voltage dividers and LED drivers.

\end{document}